\documentclass[trackchanges,astrosymb,twocolumn]{aastex701}

\newcommand{\psr}{PSR J0125\texorpdfstring{$-$}{--}5854}

\def\dmu{\ensuremath{\rm pc\,cm^{-3}}}
\def\ppdotu{\ensuremath{\rm s\,s^{-1}}}

\newcommand{\Msun}{$\mathrm{M}_{\odot} $}
\newcommand{\msun}{\mathrm{M}_{\odot}}

\newcommand{\dspsr}{{\sc dspsr}}
\newcommand{\presto}{{\sc presto}}
\newcommand{\peasoup}{{\sc peasoup}}

\newcommand{\psrchive}{{\sc psrchive}}
\newcommand{\pdmp}{{\sc pdmp}}
\newcommand{\mcmc}{MCMC}
\newcommand{\pulsarspectra}{{\sc pulsar\_spectra}}
\newcommand{\scintools}{{\sc scintools}}

\begin{document}


\title{Discovery of a 24-millisecond pulsar in a very long orbit with the Murchison Widefield Array}

\author[orcid=0000-0001-7509-0117,sname='Tan']{Chia Min Tan}
\affiliation{International Centre for Radio Astronomy Research, Curtin University, Bentley, WA 6102, Australia}
\email[show]{}  
\author[0000-0002-8383-5059]{N. D. Ramesh Bhat}
\affiliation{International Centre for Radio Astronomy Research, Curtin University, Bentley, WA 6102, Australia}
\email[show]{}  
\author[0000-0001-8845-1225]{Bradley W. Meyers}
\affiliation{Australian SKA Regional Centre (AusSRC), Curtin University, Kent Street, Bentley, WA 6102, Australia}
\affiliation{International Centre for Radio Astronomy Research, Curtin University, Bentley, WA 6102, Australia}
\email[show]{}  
\author[0000-0001-6840-4114]{Christopher P. Lee}
\affiliation{International Centre for Radio Astronomy Research, Curtin University, Bentley, WA 6102, Australia}
\email[show]{}  
\author[0000-0001-8715-9628]{Ewan D. Barr}
\affiliation{Max-Planck-Institut for Radioastronomy, Auf dem Huegel 69, D-53121 Bonn, Germany}
\email[show]{}  
\author[0000-0001-9518-9819]{Vivek Venkatraman Krishnan}
\affiliation{Max-Planck-Institut for Radioastronomy, Auf dem Huegel 69, D-53121 Bonn, Germany}
\email[show]{}  
\author[0000-0003-1307-9435]{Paulo C. C. Freire}
\affiliation{Max-Planck-Institut for Radioastronomy, Auf dem Huegel 69, D-53121 Bonn, Germany}
\email[show]{}  
\author[0000-0001-5561-1325]{Garvit Grover}
\affiliation{International Centre for Radio Astronomy Research, Curtin University, Bentley, WA 6102, Australia}
\email[show]{}  
\author[0000-0001-6114-7469]{Samuel J. McSweeney}
\affiliation{International Centre for Radio Astronomy Research, Curtin University, Bentley, WA 6102, Australia}
\affiliation{SKA Observatory, SKA-Low Science Operations Centre, Kensington, WA 6151, Australia}
\email[show]{}  
\author[0000-0001-8982-1187]{Nicholas A. Swainston}
\affiliation{International Centre for Radio Astronomy Research, Curtin University, Bentley, WA 6102, Australia}
\affiliation{CSIRO Space and Astronomy, PO Box 1130, Bentley, WA 6102, Australia}
\email[show]{}  
\author[0000-0003-0307-5633]{Qiuyang Fu}
\affiliation{National Astronomical Observatories, Chinese Academy of Sciences, 20A Datun Road, Chaoyang District, Beijing, China}
\email[show]{}  
\author[0000-0001-8018-1830]{Mengyao Xue}
\affiliation{National Astronomical Observatories, Chinese Academy of Sciences, 20A Datun Road, Chaoyang District, Beijing, China}
\email[show]{}




\begin{abstract}
We report the discovery of \psr{}, a pulsar with a spin period of 24\,ms and a dispersion measure of 11.66\,\dmu{}, in the ongoing Southern-sky MWA Rapid Two-metre (SMART) survey with the Murchison Widefield Array (MWA). The pulsar is located at a high Galactic latitude of $-57^{\circ}$, and at a distance of 0.5--1\,kpc per the Galactic electron density models. Follow-up observations with the MWA and MeerKAT telescopes have revealed that this pulsar is in a binary system with an orbital period of more than 290 days, and a steep spectrum (flux density, $ S \propto \nu^{\alpha} $, where $\nu$ is frequency and $ \alpha = -2.2 \pm 0.3 $). Analysis of current observational data hints at a potential binary configuration with an orbital period of $833.60 \pm 0.04$ days, a projected semi-major axis of $241.36 \pm 0.05$ light-seconds, and a minimum companion mass $0.4152 \pm 0.0001$\,\Msun, with a low eccentricity orbit of $0.0052 \pm 0.0006$. We discuss the potential formation channels for this system, and conjecture that the companion is likely a Helium white dwarf. Further observations are required in order to better constrain the orbital and spin parameters. We discuss the implications of this discovery, which emerged after processing a small fraction of survey data, on the prospects of finding more millisecond pulsars with the SMART survey, and with future surveys planned with the low-frequency SKA-Low.
\end{abstract}






\section{Introduction} 

Several large sky surveys and targeted searches over the past decades have led to a substantial increase in the known pulsar population, with more than 4000 pulsars currently known within our Galaxy and in the Magellanic Clouds, according to version 2.7.0 of the Australia Telescope National Facility (ATNF) pulsar catalog~\citep{mhth05}. The vast majority of these fall into two broad categories: the non-recycled or ``canonical'' pulsars, and the recycled or ``millisecond'' pulsars (MSPs). Canonical pulsars typically have spin periods ($P$) of $\mathbin{\sim} 0.1\text{--}76$\,s and spin-down rates ($\dot{P}$) of $\mathbin{\sim} 10^{-17}\text{--}10^{-12}$\,\ppdotu{}, with inferred characteristic ages ($\tau_\mathrm{c} = P / 2 \dot{P}$) of $\mathbin{\sim} 10^4\text{--}10^9$\,yr. MSPs have spin periods of $\mathbin{\sim} 1\text{--}30$\,ms and substantially smaller spin-down rates of $\mathbin{\sim} 10^{-20}\text{--}10^{-18}$\,\ppdotu, implying much larger characteristic ages of $\mathbin{\sim} 10^9\text{--}10^{10}$\,yr. MSPs are generally formed via the accretion process in a binary system where the pulsar interacts with its companion, typically a low mass star that later evolves into a white dwarf or a neutron star, resulting in the transfer of mass and angular momentum to the neutron star, leading to much shorter spin periods~\citep{tv23}.

Aside from these two broad groups, there also exists a small sub-group of pulsars with spin periods in the intermediate range $\sim$30--100\,ms, appearing in between the two major groups in the $P$-$\dot{P}$ diagram. These can be further grouped into two distinct populations based on their evolutionary characteristics. The first group tends to have rather large characteristic ages of $\mathbin{\sim} 10^8\text{--}10^9 $\,yr; they are similar to MSPs and are known as \emph{partially} (mildly) recycled pulsars. These pulsars had been through shorter accretion timescales compared to most MSPs before their companion evolved to form a second stellar remnant. Most of these are also in binary systems with their evolved stellar remnants. In most cases, the stellar companion is either a heavy white dwarf~\citep{clm+01}, or a neutron star~\citep{ht75,tkf+17}. MSPs at the higher end of the spin period distribution have shown properties similar to these partially recycled pulsars~\citep{sna+02,bdp+03}.

Identification of the binary nature of such pulsars is crucial in modeling the related stellar evolution, as the properties of the pulsar and its companion can inform the likely progenitors and the formation channels leading to the current system. Additionally, for some of these pulsars small relativistic perturbations on the orbital motion and light propagation time can be detected in the timing; these are quantified by the so-called ``post-Keplerian" (PK) parameters \citep{dt92}. By assuming general relativity, or any other theory of gravity, we can use any two PK parameters to determine the masses of the pulsar and its companion. Precise measurements of neutron star masses, especially the more massive ones \citep{fcp+21}, represent powerful constraints on the equation of state of ultra-dense matter, thus providing a better understanding of the strong nuclear force at densities above those of atomic nuclei \citep{of16}. If more than two PK parameters are measurable for the same system, then we can test general relativity and other gravity theories (\citealt{tw82,ksm+21}, see \citealt{fw24} for a review).

The second sub-group of intermediate-period pulsars have much smaller characteristic ages of $\mathbin{\sim} 10^4\text{--}10^5$\,yr, i.e. much less than that of canonical pulsars. These younger pulsars are often associated with the supernova remnants that resulted from their birth. The two notable examples are the well-known Crab pulsar, with a 33-ms spin period and the Vela pulsar with a 89-ms spin period. Such pulsars are naturally very valuable for studying the end of life of main-sequence stars and the early evolution of stellar remnants. 

Interestingly, a small number of young pulsars are found to be in binary systems with a main sequence star as companion. A notable example is PSR B1259$-$63~\citep{jml+92}, which has a spin period of 47\,ms, orbiting a Be~star companion with a binary orbital period of 1236\,days. A long-term timing study of this pulsar has provided useful insights into the formation history and spin-orbit interaction of the system~\citep{sjm14}. While having a slightly longer spin period of 132\,ms, PSR~J2032$+$4127~\citep{crr+09} is another young pulsar in a binary orbit with a Be~star companion. The binary period of this system is much longer, of the order of $\mathbin{\sim}50$\,yr~\citep{lsk+15}. X-ray observations on PSR~J2032$+$4127~\citep{hnl+17}, which were conducted during the periastron approach between the pulsar and its companion, 
have revealed the interaction between the pulsar and the companion stellar wind, providing useful hints pertaining to the evolution of such systems.

In this Letter, we report on the discovery and follow-up of \psr{}, a pulsar with a spin period of 24\,ms and a dispersion measure (DM) of 11.66\,\dmu. This is the first MSP discovered with the Murchison Widefield Array~\citep[MWA;][]{tgb+13}, and also the first pulsar found in the deep-pass (i.e. full observation) searches of the Southern-sky MWA Rapid Two-metre (SMART) survey~\citep{bsm+23a, bsm+23b}. Previously, a candidate identified by the image-plane Galactic Plane Monitor with the MWA was confirmed as an MSP by MeerKAT~\citep{phm+25}. We briefly describe details of the survey and pulsar discovery, including the follow-up observations conducted using the MWA and MeerKAT in \S\ref{sec:obs}. The results from follow-up timing and analysis are described in \S\ref{sec:results}. In \S\ref{sec:discussion}, we discuss the nature of the \psr{} system and the implications of this discovery for finding more MSPs at low frequencies, and in \S\ref{sec:conclusions}, we present our conclusions.

\section{Observations and Data Analysis} \label{sec:obs}

\subsection{Discovery in the SMART survey}

The SMART survey covers the entire southern sky at declinations below $+30 ^\circ$ with the MWA in 71 $\times$ 80-minute observations. Data are recorded as tile-level voltages with the voltage capture system~\citep[VCS;][]{tob+15,mwaxref}. These observations were taken at a central observing frequency of 154.24\,MHz with a bandwidth of 30.72\,MHz. The raw data are beamformed and then written out in the PSRFITS~\citep{hvm04} format with 100-${\mu s}$ time resolution and 3072 frequency channels across the 30.72-MHz observing band. A detailed description of the survey, including a pilot search of several 10-minute observations can be found in~\cite{bsm+23a} and some of the early discoveries and science in a series of papers~\citep{psrone,psrtwo,bsm+23b,psrfive}. A deep-pass survey has commenced in 2024, which involves searching through full 80-minute observations. Details of the search strategy and the progress with the processing to date, including new pulsar discoveries thus far, will be presented in a future publication. 

\begin{figure*}[htb] 
\centering
\includegraphics[width=.95\linewidth]{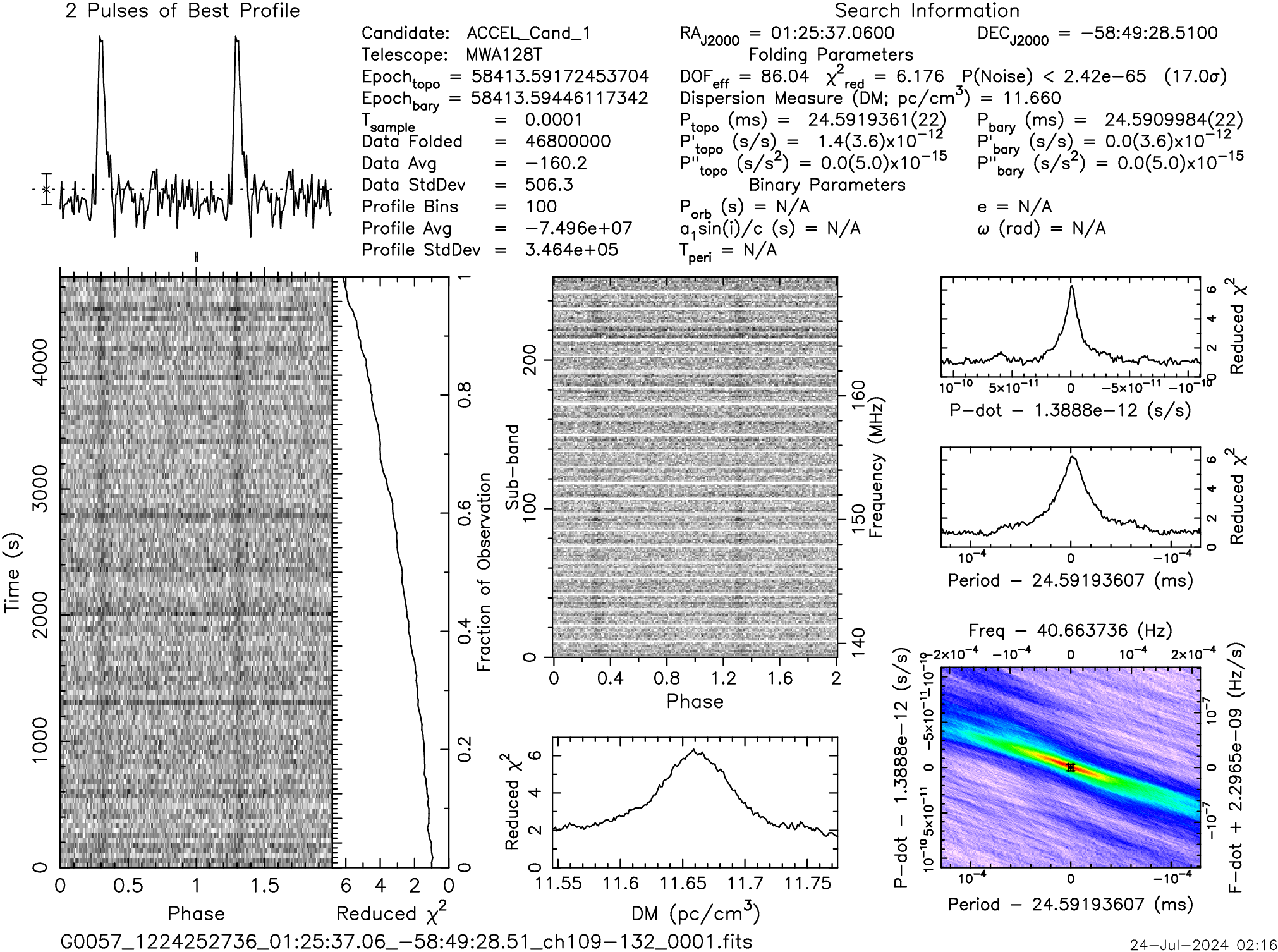}
\caption{PRESTO detection plot of \psr{} from the discovery observation: the top left panel shows the integrated (time-averaged) pulse profile; the bottom left panel displays the signal strength versus pulse phase and time; the middle panel shows signal strength versus pulse phase and frequency; the remainder plots are various diagnostic plots from the search pipeline, showing the best period and dispersion measure.}
\label{fig:discovery}
\end{figure*}

\psr{} was discovered in the SMART observation taken on 2018 October 22. The observation was centered at a position with Right Ascension and Declination of (R.A., Dec.) =  (00h01m31.05s, $-$55d06m50.28s).
The pulsar was detected in the beamformed data centered on (R.A., Dec.) = (01h25m37.06s, $-$58d49m28.51s), with a detection significance of 17$\sigma$ on the folded candidate found through the Fourier-domain search process, with a period of 24.590998(2) ms and a DM of 11.66 \dmu{}. Figure~\ref{fig:discovery} shows the diagnostic plots of detection using \presto{}.

To improve the localization, we beamformed the VCS observation with a two-ring hexagonal grid and a separation between beams of $10.13'$, which is half the full width at half maximum (FWHM) of our tied-array beam size. These data were then folded using \dspsr{}~\citep{vb11} at the discovery period and DM. The pulsar was detected on eight adjacent beams. The signal-to-noise ratio (S/N) of the detections was measured using the \pdmp{} routine from the \psrchive{} software suite~\citep{vdo12}. The position and S/N of these detections were then used to obtain an improved position for further follow-up observations. Further details of the localization procedure are described in \S\ref{sec:localisation}. 

\begin{figure*}[t] 
\centering
\includegraphics[width=1.00\linewidth]{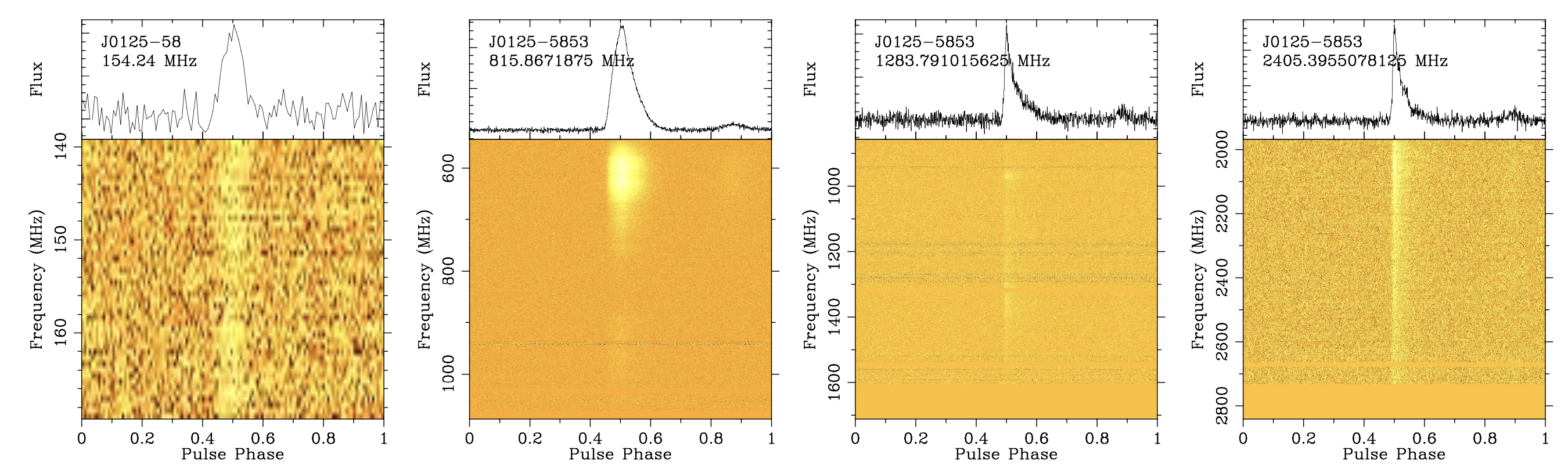}
\caption{Detection plots of~\psr{} across the frequency range spanning from $\sim$150\,MHz to 2.8\,GHz; from left to right, the MWA discovery observation (140--170\,MHz) and MeerKAT detections in the UHF, L band, and S1 band. The upper panels show the integrated profiles, and the lower panels the time-averaged signal in pulse phase vs frequency. The integration times are 80\,min for the MWA band, 60\,min for UHF, and 30\,min each for L and S1 bands. An inter-pulse emission is clearly visible in MeerKAT detections; the UHF detection is notable for an exceedingly bright scintle in the lower segment of the band.}
\label{fig:detections}
\end{figure*}

\subsection{Archival MWA data}\label{sec:mwaarchive}

A unique feature of the SMART survey design is the significant overlap ($\sim$10$^{\circ}$ in R.A. and $\sim$15$^{\circ}$ in Dec) between adjacent VCS observations, which can be exploited for independent confirmation and further characterization, as demonstrated earlier in \citet{psrone} and \citet{bsm+23b}. In addition, there are also archival data from projects that used VCS. This allowed us to retrieve observations containing \psr{}, which were then beamformed for  prospective detections. We searched through three adjacent SMART observations and detected the pulsar in one of the SMART observations, which was taken on 2018 November 23, centered at a position of (R.A., Dec.) = (00h01m20.71s, $-$72d08m46.32s). We also detected the pulsar in an archival observation (5120 seconds long) toward 47 Tucanae from 2016 November 22, centered at position of (R.A., Dec.) = (00h28m40.07s, $-$63d08m47.34s). This observation was at a higher central observing frequency of 184.96\,MHz compared to the SMART observations. All these data were folded using \dspsr{} at the discovery period and DM, and then refined using \pdmp{}. The pulsar was found to have significantly different periods between these three observations, with a change in period significantly larger than what may be expected from spin-down, suggesting that the pulsar is likely in a binary system.

\subsection{Parkes \emph{\it Murriyang} follow-up}

Following the initial confirmation of the pulsar, we made an observation using the Parkes \emph{\it Murriyang} telescope (on 2024 July 28) with the ultra-wideband low-frequency receiver~\citep{hmd+20}, with a center frequency of 2368 MHz and a bandwidth of 3328 MHz. We performed a standard grid of $5\times30$-minute observations, centered at the discovery position, with a grid of 4 observations at an offset of $11.5'$ to the North, South, East and West of the center position to cover the FWHM of the MWA tied-array beam. However, \psr{} was not detected in any of these observations. Taking into account the offset between the Parkes \emph{\it Murriyang} observation with the best-fit position in \S\ref{sec:localisation} of $4.75'$ and the modeled spectrum in \S\ref{sec:flux}, it is expected that the pulsar has an S/N of 7. As we discuss later in \S\ref{sec:discussion}, this may be reconciled in terms of the pulsar's steep spectrum and a reduced sensitivity due to radio frequency interference (RFI). 

\subsection{MWA follow-up}

Using the improved position, we initiated a dense observing campaign with the MWA, where a series of observations were taken using VCS between September 2024 and January 2025. As the MWA was operating in the long baseline configuration during this period (Phase II extended, with a maximum baseline of $\sim$5.3\,km), the $\sim 1-2'$ localization from the initial discovery (see \S\ref{sec:localisation}) proved very useful, as the FWHM of the tied array beam in this configuration is $1.1'$ at 154.24\,MHz. We successfully detected the pulsar in 9 epochs over the course of 126 days between 2024 September 10 and 2025 January 14. We also performed a single hexagonal grid around the best position to confirm the initial position. These detections are listed in Table~\ref{tab:detection_table}. The measured period was found to increase over the observing span (see Table~\ref{tab:detection_table} and Figure~\ref{fig:periods}). 

\begin{deluxetable*}{cccccccc}
\tablewidth{0pt}
\tablecaption{List of observations where \psr{} is detected. MWA P2C and P2E refers to the observation where the MWA telescope is in the compact and extended configurations respectively. $P_\mathrm{spin}$ is the measured spin period of \psr{} in the observation. $T_\mathrm{obs}$ is the total length of the observation. $T_\mathrm{res}$ is the time resolution of the observation. $\nu_\mathrm{obs}$ is the central observing frequency of the observation and $\nu_\mathrm{bw}$ is the total bandwidth of the observation. \label{tab:detection_table}}
\tablehead{
\colhead{Observation ID} & \colhead{MJD} & \colhead{Telescope} & \colhead{$P_\mathrm{spin}$ (ms)} & \colhead{$T_\mathrm{obs}$ (s)} & \colhead{$T_\mathrm{res}$ ($\mu$s)} & \colhead{$\nu_\mathrm{obs}$ (MHz)} & \colhead{$\nu_\mathrm{bw}$ (MHz)}
}
\startdata
1163853320 & 57714.555288 & MWA P2C & 24.5910444(15) & 5120 & 100 & 184.96 & 30.72\\
1224252736 & 58413.622158 & MWA P2C & 24.5909866(15) & 4800 & 100 & 154.24 & 30.72\\
1227009976 & 58445.533206 & MWA P2C & 24.5910418(15) & 4800 & 100 & 154.24 & 30.72\\
1410025688 & 60563.766047 & MWA P2E & 24.590056(6) & 3600 & 100 & 154.24 & 30.72\\
1414506224 & 60615.622691 & MWA P2E & 24.590063(5) & 3600 & 100 & 154.24 & 30.72\\
1415109368 & 60622.603208 & MWA P2E & 24.590081(2) & 3600 & 100 & 154.24 & 30.72\\
1415712520 & 60629.583800 & MWA P2E & 24.590085(2) & 3600 & 100 & 154.24 & 30.72\\
1416315672 & 60636.564379 & MWA P2E & 24.590107(4) & 3600 & 100 & 154.24 & 30.72\\
1416918816 & 60643.544857 & MWA P2E & 24.590113(3) & 3600 & 100 & 154.24 & 30.72\\
1417263472 & 60647.533726 & MWA P2E & 24.590115(4) & 3600 & 100 & 154.24 & 30.72\\
1417521968 & 60650.525425 & MWA P2E & 24.590126(2) & 3600 & 100 & 154.24 & 30.72\\
1420890072 & 60689.506312 & MWA P2E & 24.590242(3) & 3600 & 100 & 154.24 & 30.72\\
1741615626 & 60744.610938 & MeerKAT & 24.590411(2) & 3600 & 120 & 815.867 & 544\\
1746952274 & 60806.367967 & MeerKAT & 24.590650(2) & 1800 & 77 & 1283.791 & 856\\
1750997776 & 60853.193148 & MeerKAT & 24.590820(4) & 1800 & 19 & 2405.396 & 876\\
\enddata
\end{deluxetable*}

\subsection{MeerKAT follow-up}

Following the non-detection with \emph{Murriyang} and the non-availability of the MWA for further observations with VCS (starting 2025 February), we initiated observations with the MeerKAT telescope through a series of Director's Discretionary Time projects. In total, three observations were made between March and June 2025, the details of which are given in Table~\ref{tab:detection_table}. The first observation was conducted on 2025 March 10 in the UHF band, with a central frequency of 815.867\,MHz and a bandwidth of 544 MHz over 2048 channels, and a total integration time of 60 minutes. This used 60 MeerKAT dishes, and the data was recorded by the Accelerated Pulsar Search User Supplied Equipment (APSUSE) backend~\citep{cbk+21,pbs+23}, after beamforming using the Filterbanking Beamformer User Supplied Equipment (FBFUSE), in filterbank mode with a sampling time of 120.471\,$\mu$s. A total of 477 beams were recorded, centered at the best known position from the localization analysis with the MWA (see \S\ref{sec:localisation}), with the beams overlapping at 75\% of maximum sensitivity, covering a circular region of radius $2.1'$. Data from all these beams were first searched using \peasoup{}~\citep{mbc+19,pbs+23} to obtain an initial position, and those with high S/N were subsequently folded at the best period and DM of the pulsar from the MWA observations. The pulsar was clearly detected in a large fraction of these beams, with S/N ranging from $\sim$100 to $\sim$1000. As seen in Figure~\ref{fig:detections}, this extremely bright detection is caused by the fortuitous occurrence of an ultra-bright \emph{scintle} in this observation. 

A second observation was conducted on 2025 May 11 using the L-band receiver. This one used 52 dishes for beamforming, with a center frequency of 1283.791\,MHz, and a bandwidth of 856\,MHz, split into 2048 channels, after downsampling by a factor of two after detection. A shorter 30-minute observation was recorded using APSUSE in  filterbank mode (after beamforming by FBFUSE), with a sampling time of 76.561\,$\mu$s. This time a smaller grid of 27 beams was formed, centered at the best position obtained from the localization analysis with UHF data. The beams were tiled with an overlap at 75\% of the maximum sensitivity, and the data were folded at the best period and DM from the MWA observations. The pulsar was detected in 22 adjacent beams, with a maximum S/N of $\sim$80.

A third observation was conducted on 2025 June 27, using the S-band receiver, with a center frequency of 2405.396\,MHz and a bandwidth of 
875\,MHz (S1 band). This observation used 56 dishes, and after performing a downsampling by factor of two, the native 1024 channel data were written as 512 channels across the band. A 30-minute observation was recorded using the APSUSE backend (after beamforming by FBFUSE) in the filterbank mode with a sampling time of 18.725\,$\mu$s. As with the L-band observation, 27 beams were formed around the best position obtained from the localization using L-band data, with a beam separation at 75\% of maximum sensitivity. The data from these beams were folded at the best period and DM from the MWA observations and the pulsar was detected in 27 adjacent beams, with a maximum S/N of $\sim$100. As seen in Figure~\ref{fig:detections}, the S1 band detection is visibly brighter compared to that in the L band, which may be attributed to another episode of fortuitous scintillation brightening. 

\begin{figure}[t]
\centering
\includegraphics[width=\linewidth]{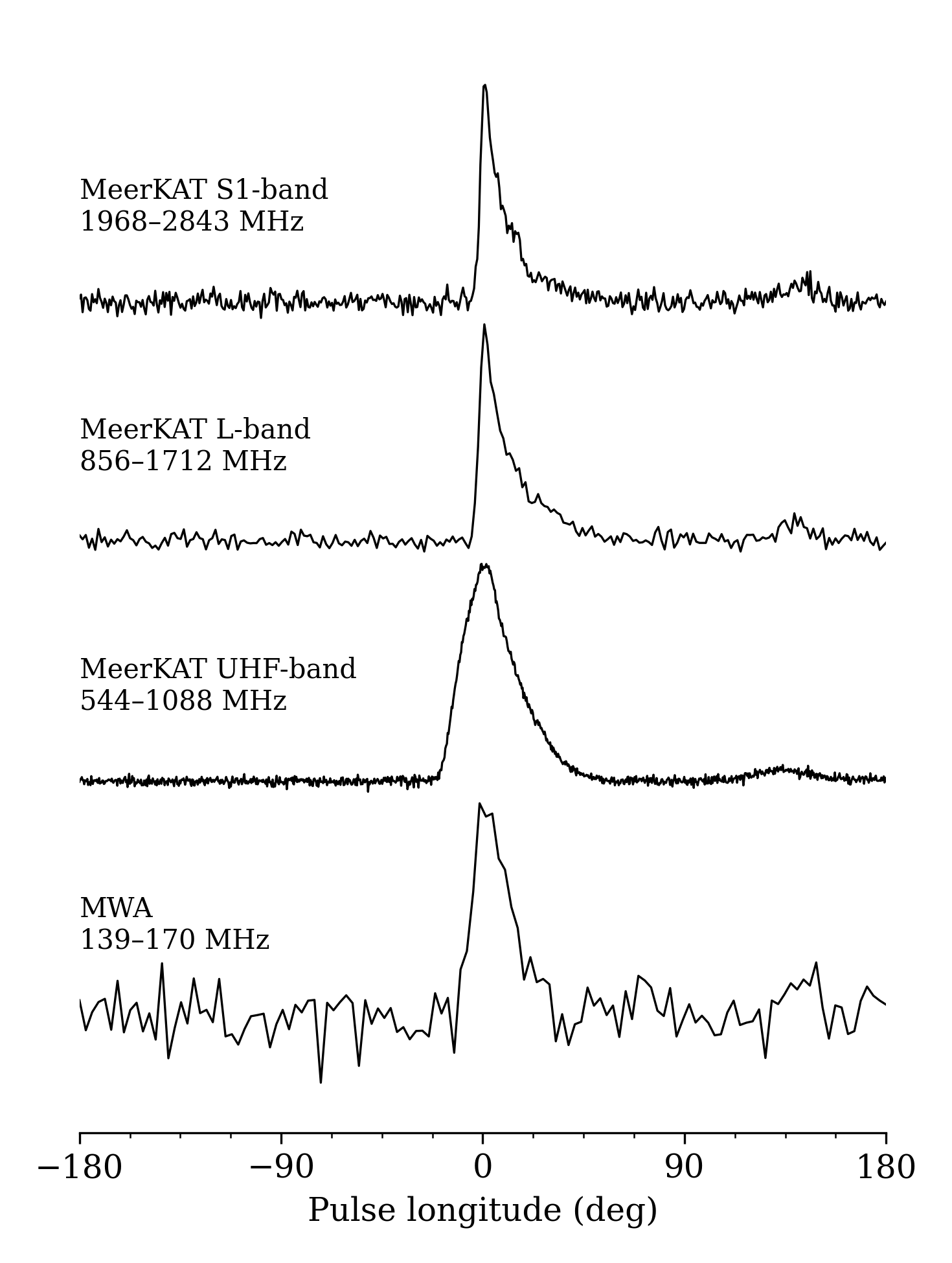}
\caption{Integrated profiles of \psr{} from MWA and MeerKAT observations, showing the profile evolution across the frequency range from $\sim$100 MHz to 3\,GHz. The MWA profile is from an 80 min integration, whereas for MeerKAT, it is 60 min in the UHF band, and 30 min each in the L and S1 bands. MeerKAT profiles are at a higher time resolution (1024 phase bins) compared to the MWA profile (128 phase bins). 
}
\label{fig:profiles}
\end{figure}

Figure~\ref{fig:profiles} shows integrated profiles across the MWA and MeerKAT frequency bands, spanning $\sim$150\,MHz to $\sim$3\,GHz. The MeerKAT profiles are at higher time resolutions, with 1024 phase bins across the pulse period, whereas the MWA profile has a much coarser time resolution (128 phase bins). The MeerKAT profiles also show clear hints of an inter pulse, at $\sim135^{\circ}$ offset from the profile peak. There is some intrinsic broadening at lower frequencies, but in light of the poorer resolution and lower S/N of the MWA profile, we refrain from commenting on plausible physical interpretations. 

\section{Analysis and Results} \label{sec:results}

\begin{deluxetable}{lc}
\tablewidth{0pt}
\tablecaption{Parameter summary of \psr{} \label{tab:property}}
\tablehead{
\colhead{Parameter} & \colhead{Value} 
}
\startdata
Right Ascension & 01h25m47.31s\\
Declination & $-$58d54m02.24s\\
Galactic Longitude ($^{\circ}$) & 294.64\\
Galactic Latitude ($^{\circ}$) & $-$57.67\\
Discovery spin period (ms) & 24.590998(2)\\
Dispersion Measure (\dmu) & 11.663(4)\\
DM distance (NE2001) (kpc)  & 0.54\\
DM distance (NE2025) (kpc)  & 0.75\\
DM distance (YMW16) (kpc) & 1.03\\
Flux density at 154 MHz, $S_{154}$ (mJy) & 17(9)\\
Flux density at 1284 MHz, $S_{1284}$ (mJy) & 0.18(9)\\
Spectral index ($\alpha$) & $ -2.2(3) $\\
\enddata
\end{deluxetable}

\subsection{Localization} \label{sec:localisation}

\begin{figure}[t] 
\centering
\includegraphics[width=\linewidth]{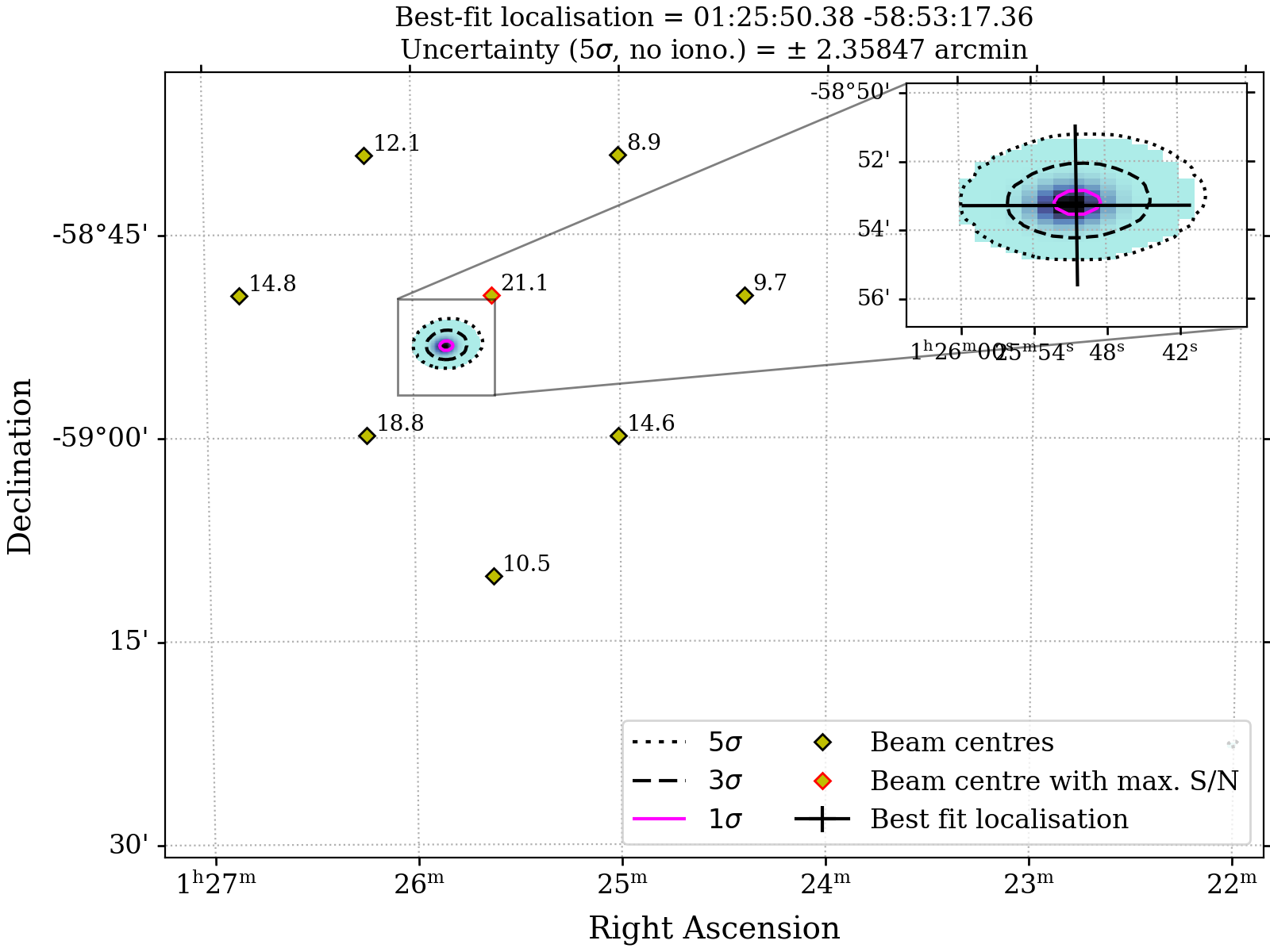}
\caption{The MWA localization of \psr{} from the discovery observation.
The nine adjacent detections were able to constrain the pulsar position to within a $2.4^\prime$ (5$\sigma$) confidence region, shown by the dotted ellipse. }
\label{fig:localisation}
\end{figure}

The localization method that uses MWA observations~\citep{mb26} employs a maximum-likelihood estimation approach that is similar to the method implemented in SeeKAT~\citep{bcb+23} but suitably adapted for MWA, taking into account the complexity of the array layout and beam response. In short, it leverages the detection significances in a densely sampled grid of positions and the knowledge of tied-array beam shape. Application of this method yields a pulsar position (R.A., Dec.) = (01h25m50s, $-$58d53m17s), with a $5\sigma$ position error of $2.4^\prime$ (${\sim}0.04^\circ$). However, this does not account for any potential ionospheric refractive shift. The localization region is shown in Figure~\ref{fig:localisation}, and this was used as the starting footprint for the follow-up MeerKAT observation in the UHF band.

\begin{figure*}[t] 
\centering
\includegraphics[width=\linewidth]{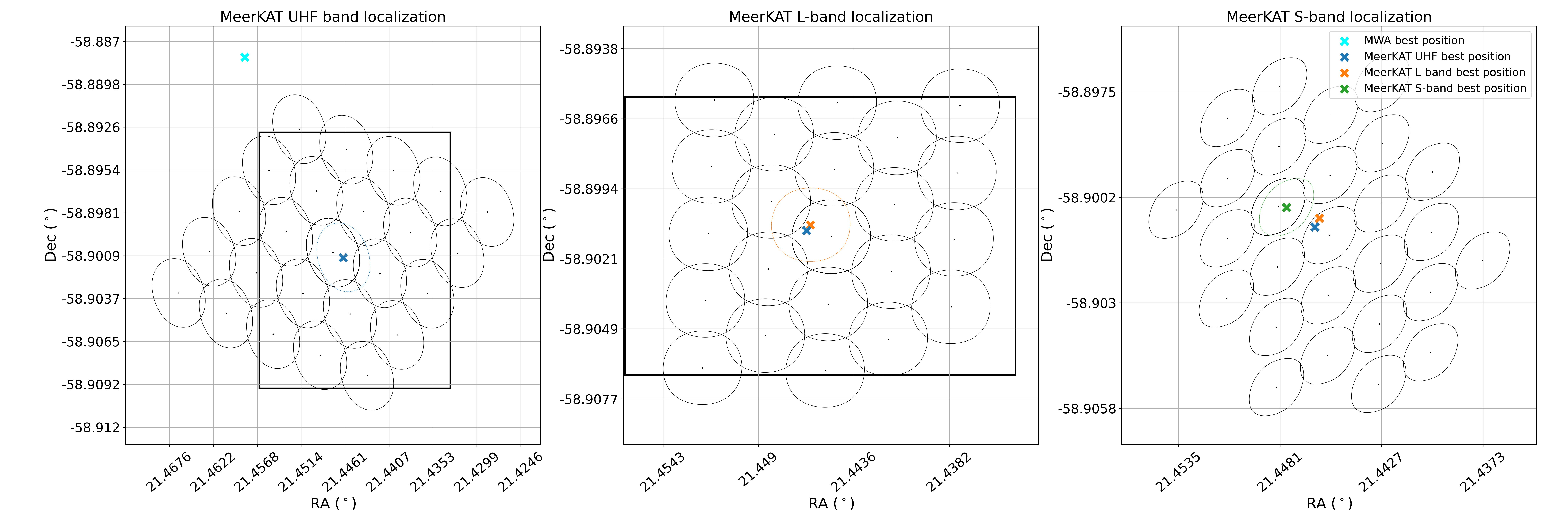}
\caption{The localization plots obtained from SeeKAT on the three MeerKAT observations. The left plot is the localization position obtained from the UHF-band observation. The middle plot is the localization position obtained from the L-band observation and the right plot is the localization position obtained from the S1-band observation. The grid on each plot are drawn at intervals of 10 arcseconds. The boxes in the first 2 plots are the zoomed-in region for the subsequent localization at higher frequency.
}
\label{fig:meerkat_localisation}
\end{figure*}

The MeerKAT UHF observation covered an area about twice the uncertainty region of the initial MWA localization. The SeeKAT package was used for the localization analysis of the observation. We first simulated a point spread function (PSF) of the response of the MeerKAT array used with 4000 pixels and with a resolution of $1.0''$ using Mosaic\footnote{\url{https://github.com/wchenastro/Mosaic}}~\citep{cbk+21}. Using this PSF and S/N of the 8 brightest detections, the best-fit position was determined to be (R.A., Dec.) = (01h25m46.76s, $-$58d54m04.60s)\footnote{Note that the UHF band localization also resulted in a slight revision of the pulsar name.}. However, we were not able to obtain reliable estimates of the uncertainty in the localization. The localization plot of the UHF observation is shown in the left panel of Fig.~\ref{fig:meerkat_localisation}. The UHF localized position is $\approx$1.21$^{\prime}$ offset from the MWA-determined position, thus providing an important (and independent) validation of our initial method. The localization was further refined using the S/N of 25 detections, where we simulated a finer PSF, with 2.56 million pixels and  a resolution of 0.04$^{\prime\prime}$. This yielded a best-fit position of (R.A., Dec.) = (01h25m47.08s, $-$58d54m03.88s). 

For localization using L-band observation, we adopted a similar approach by simulating a PSF with 2.56 million pixels and a resolution of 0.03$^{\prime\prime}$, and used the S/N of 22 detections in the L-band observation. The best-fit position using SeeKAT is (R.A., Dec.) = (01h25m47.05s, $-$58d54m02.90s). For localization using S1 band observation, we simulated a PSF with 1.44 million pixels and a resolution of 0.02$^{\prime\prime}$. Here we used the S/N of all 27 detections and  the best-fit position (using SeeKAT) is (R.A., Dec.) = (01h25m47.06s, $-$58d54m01.88s). The localization plots of the L and S1 band observations are shown in the middle and right panels of Fig~\ref{fig:meerkat_localisation}, respectively.

Throughout the localization process using SeeKAT, we estimated the localization uncertainty as the width of the tied-array beam at the overlapping level in each observation and used that to compute the weighted best-fit position from the three MeerKAT observations. The best-fit position determined in this manner is (RA, Dec) = (01h25m47.31s, $-$58d54m02.24s), with an uncertainty of $2.1^{\prime\prime}$ in Right Ascension and $2.3^{\prime\prime}$ in Declination. This final best-fit position is approximately $49^{\prime\prime}$ offset from the original MWA position and thus matches well within the initial uncertainty region. A summary of all localization analyses is presented in Table~\ref{tab:localisation}.

\begin{deluxetable}{ccccc}
\tablewidth{0pt}
\tablecaption{A summary of the localization results from the different observations of \psr{}. \label{tab:localisation}}
\tablehead{
\colhead{Telescope} & \colhead{$\nu_\mathrm{obs}$ (MHz)} & \colhead{R.A.} & \colhead{Dec} & \colhead{Distance from best position}
}
\startdata
MWA P2C & 154.24 & 01h25m50s & $-$58d53m17s & 49.8$^{\prime\prime}$\\
MeerKAT & 815.867 & 01h25m47.08s & $-$58d54m03.88s & 2.42$^{\prime\prime}$\\
MeerKAT & 1283.791 & 01h25m47.05s & $-$58d54m02.90s & 2.12$^{\prime\prime}$\\
MeerKAT & 2405.396 & 01h25m47.46s & $-$58d54m01.88s & 1.26$^{\prime\prime}$\\
\hline
Best position & & 01h25m47.31s & $-$58d54m02.24s & -- \\
\enddata
\end{deluxetable}

\psr{} is located at a high Galactic latitude of $|b| = 57.67^{\circ}$, with an estimated distance of 0.54 kpc per the NE2001 model~\citep{cl02}, 1.03 kpc per the YMW16 model~\citep{ymw16} and 0.75 kpc per the NE2025 model~\citep{oc2026}. This implies a scale height of 0.3--0.55\,kpc. The discrepant distances at these high latitudes point to the need for further refinements of the halo component of the electron density models \citep{occ2021}, which are particularly useful in interpreting the measurements of fast radio bursts (FRBs) detected at these high latitudes. 

\subsection{Orbital analysis} \label{sec:orbit}

\begin{figure*}[ht!] 
\includegraphics[width=.95\linewidth]{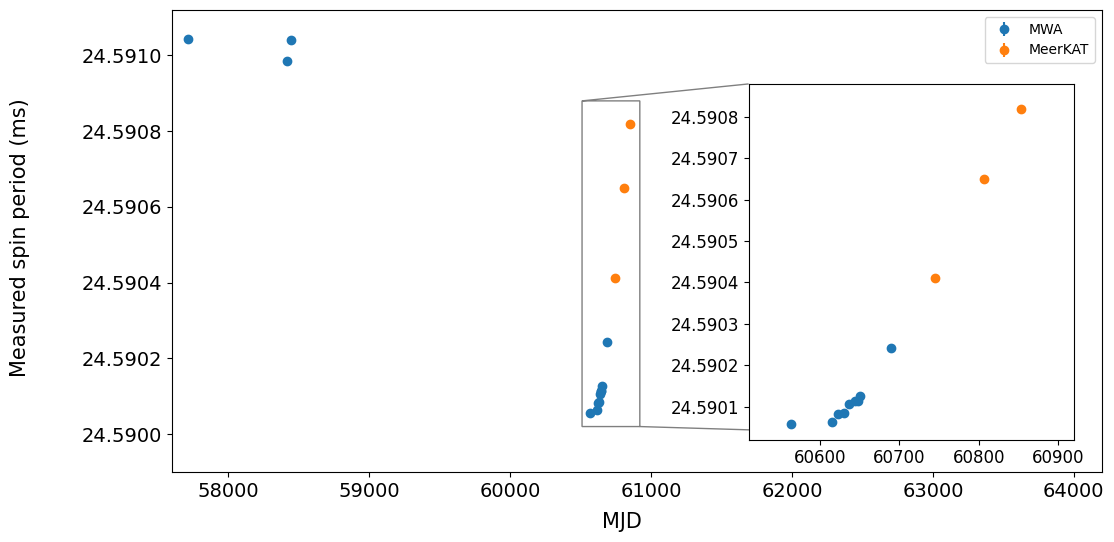}
\caption{The measured barycentric spin period of \psr{} from all the observations where the pulsar is detected. Inset : Zoomed in plot of the measured spin periods from MJD 60550. Blue points are observations from the MWA telescope and orange points are observations from the MeerKAT telescope. The error-bars in the measured spin period are small relative to the scale of the plot. The scatter in the measured spin periods from the MWA observations between MJD 60563--60689 are likely due to low S/N of the detections resulting in measurement uncertainties.
}
\label{fig:periods}
\end{figure*}

For all MeerKAT observations where the pulsar was detected, we folded the beam data with the highest S/N using \dspsr{} and a simple ephemeris containing the best period obtained from our initial analysis, to produce timer archive files. We then used the \pdmp{} routine of \psrchive{} to further refine the spin period at each observing epoch. Fig.~\ref{fig:periods} and Table~\ref{tab:detection_table} show the barycentric spin periods measured from all these observations, with three observations from MJD 57714--58445 and the other 12 from MJD 60563--60853, with a gap of almost 6 years between the two sets of observations.

As discussed in \S\ref{sec:mwaarchive}, the three initial observations (with the MWA), including the discovery observation from SMART suggested that the spin period of the pulsar changes significantly over time, with a trend that is not consistent with a pulsar spinning down. Observations taken after MJD 60563 unambiguously confirm this, where the measured spin period was found to be smaller than that in observations taken before MJD 58445. This suggested that the variation in the spin period is most likely due to Doppler shifts from the orbital motion around a binary companion.

The observed changes in the spin period between MJD 60563--60853 suggested that the binary orbit of the pulsar is larger than the observing span of 290 days. For a binary pulsar that has completed a full orbit, we expect to observe a cyclic variation in the observed spin period over time, with two turnovers at the minimum and maximum observed spin period values. Here we only observed a hint of turnover in the observed spin period at MJD $\sim$ 60600, while the maximum observed spin period is still smaller than the ones seen at MJD 57714--58445, implying that \psr{} has not completed a full revolution around its companion.

Initial attempts to model the variation of this apparent period over the full 8 year time span were unsuccessful, due to the large gap between recent follow-up observations and the three earlier observations. Later, We were able to constrain some of the orbital parameters using a Markov Chain Monte Carlo (MCMC) fit to only period measurements obtained between MJD 60563--60853, assuming a Keplerian orbit. The initial orbital parameters were then further refined through modeling of the times-of-arrivals of \psr{} from all observing epochs.

\subsubsection{MCMC fit to observations between MJD 60563--60853}

\begin{figure}[ht!] 
\includegraphics[width=.95\linewidth]{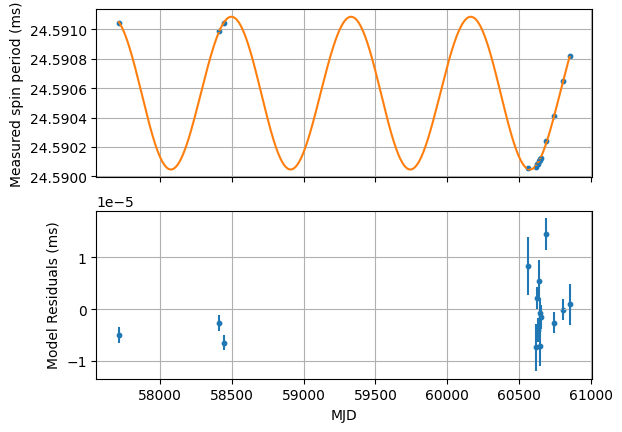}
\caption{Top: The solid line shows the variation of the spin period expected from the best-fit orbital model for the~\psr{} system and the blue dots showing the measured period of the pulsar at all observation epochs. Bottom: The residuals in the measured spin periods compared to the expected values from the modeled orbit. The rms value and the reduced $\chi^2$ of the fit $5 \times 10^{-6}$ ms and $7.3$ respectively.}
\label{fig:best_orbit}
\end{figure}

We first made initial guesses for the potential orbit using the results from {\tt fit\_orbit.py} routine from {\sc PSRpy}\footnote{\url{https://github.com/emmanuelfonseca/PSRpy}}, which is then used as a starting point to run multiple~\mcmc{} fits, while fixing the eccentricity to a range of values from $10^{-5}$ to $0.9$. Further details on the MCMC analysis is provided in Appendix~\ref{sec:appendix}. We were able to obtain constraints on the orbital properties for low eccentricities of $ e \lesssim 0.05$; however, we were unable to obtain any meaningful constraints for larger values. For example, at $e = 0.01$, the 68\% confidence levels of the binary period and projected semi major axis of the system are $ P_b = 833^{+29}_{-31}$ days and $ x = 240^{+16}_{-18}$ light-seconds respectively, with a minimum companion mass of $ M_{\rm c,min} = 0.41 \pm 0.02$ M$_\odot$. To verify if the orbital constraints are reasonable, we used the modeled orbital properties to predict the expected spin period of the pulsar at MJDs 57714, 58413, and 58445 and found that the 68\% confidence levels are $24.59098^{+0.00007}_{-0.00025}$\,ms, $24.59098^{+0.00011}_{-0.00026}$\,ms and $24.59103^{+0.00006}_{-0.00020}$\,ms respectively. All of these predicted spin periods are in agreement with the measured spin period of \psr{} at their respective epochs, giving strong credence to a low eccentricity binary system. One of the best-fit orbital models from the MCMC fit at a period of 833 days is shown in Figure~\ref{fig:best_orbit}, showing that the modeled orbital parameters fits very well with the measured periods from the earlier epochs, with an overall rms value of $5 \times 10^{-6}$ ms and a reduced $\chi^2$ of $7.3$.

\subsubsection{Solution refinement through TOA modeling}

We further refined the orbital parameters of~\psr{} by modeling the times-of-arrival (TOAs) of the pulsar at each observing epoch. We obtained three TOAs for each MWA observation, 12 TOAs from the MeerKAT UHF observation, and 6 TOAs each from the other MeerKAT observations. In this method, we used the best-fit orbital model from the MCMC fit as an initial guess of the orbit, added arbitrary jumps between each observing epoch, assumed that the pulsar position to be at the best-fit position from the localization effort, and that there is no significant spin period derivative. The spin period and the orbital parameters of~\psr{} are then iteratively fitted with \textsc{tempo}, with a bootstrap method to estimate the uncertainties in the parameters. Using this method, we obtained $ P_b = 833.60 \pm 0.04$ days, $ x = 241.36 \pm 0.05$ light seconds, and $e = 0.0052 \pm 0.0006$; these values represent a three orders of magnitude improvement in precision relative to the spin period analysis. This suggests $ M_{\rm c,min} = 0.4152 \pm 0.0001$ M$_\odot$.

\begin{deluxetable}{lc}
\tablewidth{0pt}
\tablecaption{Orbital parameter summary of \psr{} \label{tab:orbital_property}}
\tablehead{
\colhead{Parameter} & \colhead{Value} 
}
\startdata
Binary period, $P_b$ (days) & 833.60(4)\\
Projected Semi-major Axis, $x$, (lt-s) & 241.36(5)\\
Epoch of Ascending Node, $T_{asc}$ (MJD) & 60163.12(13)\\
First Laplace-Lagrange parameter, $\epsilon_{1}$ & 0.0024(2)\\
Second Laplace-Lagrange parameter, $\epsilon_{2}$ & -0.0046(7)\\
\hline
Eccentricity, $e$ & 0.0052(6)\\
Longitude of periastron, $\omega$ ($^{\circ}$) & 152(4)\\
Time of periastron, $T_0$ (MJD) & 60516(9)\\
Minimum companion mass, $M_{\rm c,min}$ & 0.4152(1)\\
\enddata
\end{deluxetable}

While the refined orbital parameters are obtained through pulsar timing, we are yet to obtain a coherent timing solution for the pulsar as this method did not preserve phase coherence between observing epochs, which means that we could not yet fit for the position and the spin down of the pulsar. Using the {\sc dracula}\footnote{\url{https://github.com/pfreire163/Dracula}} package~\citep{2018MNRAS.476.4794F}, we still obtain many possible phase-coherent timing solutions for the whole data set with positions in agreement with the preliminary position derived in this work. The family of solutions are all consistent with the uncertainty in the parameters presented in Table~\ref{tab:orbital_property}. Additional data will be needed to find the correct one.

\subsection{Flux densities and Spectrum}\label{sec:flux}

We measured the pulsar flux densities using observations from MWA and MeerKAT in order to constrain its spectrum. The flux densities at $\sim$100--200\,MHz were measured using data from the discovery observation (140--170\,MHz) and archival data from a past observation (170--200\,MHz) of the 47-Tucanae globular cluster. As the pulsar was detected mainly in the lower half of the observing band of the 47-Tucanae observation (likely due to poor calibration of the upper part of the observing band), the flux density measurement was made in this segment of the observing band, i.e. at a central frequency of 177.28\,MHz and a bandwidth of 15.36\,MHz. The flux densities measured from the MWA observations were computed by simulating the system-equivalent flux density (SEFD) of the telescope using the method described and implemented by~\cite{lbm+25} that builds on the original framework by~\cite{mtb+17}. Simulations were performed at four different frequency bands and across four different observing sub-integrations of each observation, and the mean SEFD over these simulations was used for flux calibration. 

The flux densities in the three MeerKAT observations were estimated using the radiometer equation, as it has been demonstrated to be a reliable approach for MeerKAT flux measurements~\citep{2022PASA...39...27S,gitika2023}. In order to obtain multiple measurements across the large MeerKAT band, with considerations on the offset for instrumental spectral index, each MeerKAT observation was split into two sub-bands. The SEFD\footnote{SEFD values of the MeerKAT receivers are taken from \url{https://skaafrica.atlassian.net/wiki/spaces/ESDKB/pages/277315585/MeerKAT+specifications}} of the UHF receiver was taken to be 550\,Jy at 679.867\,MHz and 400\,Jy at 951.867\,MHz respectively, while the SEFD of the L-band receiver was taken to be 450\,Jy at 1069.791\,MHz and 395\,Jy at 1497.791\,MHz respectively. The SEFD of the S1-band receiver was taken to be 460\,Jy at 2186.646\,MHz and 450\,Jy at 2616.146\,MHz respectively. We also corrected for the difference in the beam response due to the offset between the position of the beams where the detection is made compared to the best fit position obtained in \S\ref{sec:localisation}. This resulted in correction factors of 0.993, 0.987, 0.821, 0.710, 0.934 and 0.908 applied to the observations at observing frequencies of 679.867\,MHz, 951.867\,MHz, 1069.791\,MHz, 1497.791\,MHz, 2186.646\,MHz and 2616.146\,MHz respectively. As these corrections are calculated only using the central frequency of a large bandwidth, and the best-fit localization position has an uncertainty of $\sim 2''$, we quoted the uncertainty of the flux densities measured to be 50\% of their values.

\begin{figure}[t] 
\includegraphics[width=.95\linewidth]{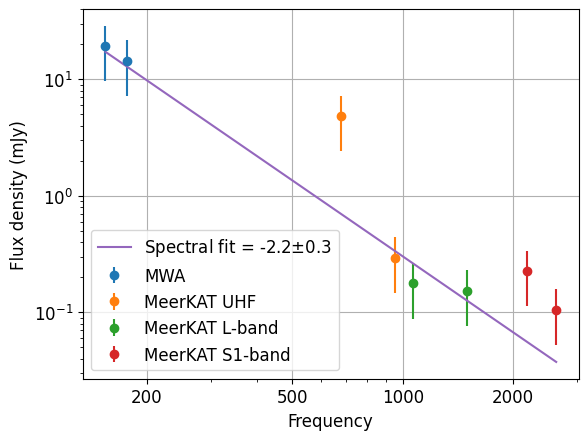}
\caption{The spectrum of \psr{} from MWA and MeerKAT detections. 
The best-fit line is from an unbiased fit to all measurements shown, including the one near 600\,MHz (the UHF band) and those above 2\,GHz (S1 band) that are likely scintillation boosted.}
\label{fig:spectrum}
\end{figure}

The overall pulsar spectrum was modeled as a simple power-law, where mean flux density $ S\propto\nu^{\alpha} $ (and $\nu$ is frequency), across the $\sim$100\,MHz to $\sim$3\,GHz observing frequency range, is shown in Fig.~\ref{fig:spectrum}. We used the spectral fitting method implemented within \pulsarspectra{}~\citep{2022PASA...39...56S}, which employs the Huber loss function to down-weight outliers in the fit. This mitigates the bias on the fit due to the ultra-bright scintle in the UHF band (see Figure~\ref{fig:detections}), and possible scintillation in the S1 band. For our current data, this yields a spectral index, $\alpha = -2.2 \pm 0.3$. If the S1 band flux densities are biased high due to scintillation, as they appear in the plot, then the true spectrum may be steeper than our estimate. Longer term flux-density monitoring will be necessary to further constrain the spectrum. 

We have attempted to measure some basic scintillation properties of \psr{} from the UHF observation using the autocorrelation function of the \scintools{} package~\citep{rcb+20,rc23}. This yields a scintillation bandwidth $ \Delta \nu_d \sim$ 60\,MHz; however, we were unable to measure the scintillation timescale, as it is more than the length of our observation (i.e. $ \Delta t _{\rm iss} > 1 $\,hour). Moreover, due to the presence of a single (fraction of) a scintle in this observation, uncertainties can be large, as much as 100\%. Nevertheless, based on these initial measurements, and assuming the NE2001 distance of 0.54\,kpc, we estimate a mean turbulence strength of $\overline { C_n^2 } \sim 10^{-4} \, {\rm m^{-20/3} }$ and an implied scintillation velocity $ V_{\rm iss} < $ 75\,${\rm km\,s^{-1}} $. This value of $\overline { C_n^2 }$ is comparable to that estimated toward 
the MSP PSR J0437$-$4715 \citep{2014ApJ...791L..32B}, and thus one of the lowest of all measurements published so far. 

\section{Discussion} \label{sec:discussion}

\subsection{Nature of the binary system}\label{sec:nature}

\begin{figure}[ht!] 
\includegraphics[width=.95\linewidth]{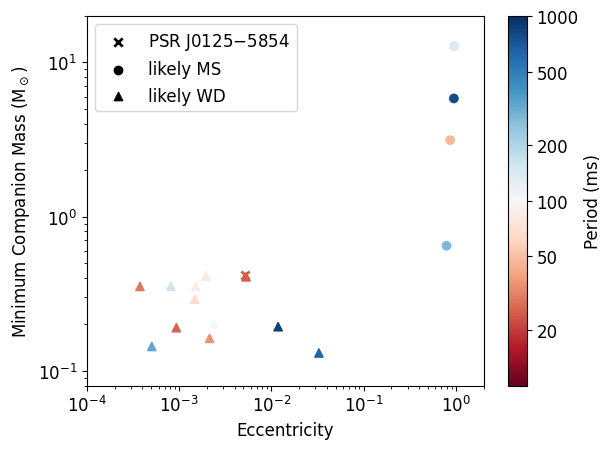}
\caption{Scatter plot showing the eccentricity and the minimum companion mass of the 16 pulsars with known binary orbital period of more than 290 days. The circle points are companions that are most likely to be main sequence stars and the triangle points are companions that are most likely to be white dwarfs. \psr{} is shown as a cross. The color of the points shows the spin period of the pulsars in their respective binary system.}
\label{fig:ecc_minmass}
\end{figure}

Binary pulsars with orbital periods similar to \psr{} are rare within our Galaxy. In fact, according to version 2.7.0 of the ATNF pulsar catalog~\citep{mhth05}, there are only 20 systems with an orbital period larger than 290 days, out of 571 known binary systems. Figure~\ref{fig:ecc_minmass} shows a summary of 16 of these systems that have a measured orbital eccentricity, $e$, where it is plotted against their minimum companion mass, $M_{\rm c,min}$. As seen from this figure, there appear to be two distinct populations.

The first group of 12 pulsars are in nearly circular binary orbits, with low eccentricities ($e \lesssim 0.05 $), and relatively low mass companions. The estimated minimum masses for these systems range from 0.13--0.41\,$\textrm{M}_\odot$. 
These systems are thought to be pulsar--helium white dwarf (He WD) systems. Generally, pulsars with He WD companions have spin periods of a few milliseconds. However, for orbital periods longer than 200 days, all of them have spin periods of tens to hundreds of milliseconds. All pulsar--He WD systems form in case-B Roche-lobe overflow, which happens after hydrogen burning ceases in the WD progenitor, and the star enters the giant branch \citep{tv23}. However, for wider systems, the mass transfer phase is much shorter, occurring only when the system is near the tip of the giant branch. For this reason, they are not as strongly recycled as the pulsars in the shorter systems.

The second group of 4 pulsars has much larger eccentricities of $e \gtrsim 0.8$ and relatively larger minimum companion masses of $0.64\, \textrm{M}_\odot \lesssim \rm M_{\rm c,min} \lesssim 12.76\, \textrm{M}_\odot $. Two of these pulsars, PSRs B1259$-$63~\citep{jml+94} and J2032$+$4127~\citep{lsk+15}, are known to harbor main-sequence star companions of type Be. The companions of the other two pulsars, PSRs B1820$-$11~\citep{cl86, hlk+04} and J1638$-$4725~\citep{lfl+06}, have not yet been identified, although both are speculated to be main-sequence stars~\citep{pv91, lfl+06}. Although not listed in the ATNF pulsar catalog, PSR J0210$+$5845~\citep{vbj+24} is known to be in a $47^{+40}_{-14}$ years long binary orbit with a high eccentricity $e=0.46^{+0.10}_{-0.07}$, and a main-sequence companion of spectral type B6 V, with an estimated mass of $\rm M_{\rm c} \sim 3.5\text{--}3.8\,\mathrm{M}_\odot$. Importantly, all five have rotational properties that are consistent with non-recycled pulsars, which suggests that these binary systems are yet to undergo an accretion process. Considering the large orbital separation, the accretion process is most likely only going to begin when the main sequence companion has further evolved.

The orbital model of \psr{} obtained from \S\ref{sec:orbit} suggests a low observed eccentricity, which indicates that this binary system is in the first group of systems with wide orbits, implying that the companion is a Helium white dwarf star. The properties of \psr{} are very similar to two of these pulsars, PSRs J0214$+$5222~\citep[$P \sim 24.6$ ms, $P_b \sim 512$ days][]{flm+23} and J0407+160~\citep[$P \sim 25.7$ ms, $P_b \sim 669$ days][]{lxf+05}. Other pulsars with slightly longer spin periods in the 30--100 ms range include PSRs J1840$-$0643~\citep{kek+13}, J2016$+$1948~\citep{naf03} and J2204$+$2700~\citep{mgf+19}, as well as the globular cluster pulsars PSRs J1953$+$1846B and J1953$+$1846C~\citep{lpz+25}

According to \cite{TS99}, the mass of a He WD is directly related to the orbital period as: 
\begin{equation}
\frac{M_\mathrm{WD}}{\msun} = \left( \frac{P_b}{b} \right)^{1/a} + c,
\end{equation}
where the values of $a$, $b$ and $c$ are given, for different types of progenitors, by:
\begin{itemize}
\item 4.50, $1.2 \, \times 10^5$, 0.120 for Population I,
\item 4.75, $1.1 \, \times 10^5$, 0.115 for Population I+II,
\item 5.00, $1.0 \, \times 10^5$, 0.110 for Population II.
\end{itemize}

\begin{figure}[ht!] 
\includegraphics[width=.95\linewidth]{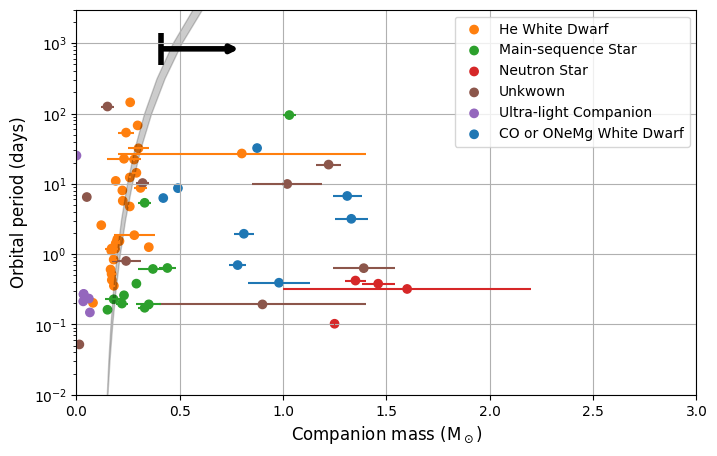}
\caption{Scatter plot showing the relationship between the orbital period of a binary pulsar system and the mass of the companion, for different types of companion, all with known masses, obtained from version 2.7.0 of the Australia Telescope National Facility (ATNF) pulsar catalog. The shaded region is the expected relationship between the orbital period and the mass of a He WD for the different types of progenitors as described in \cite{TS99}. The minimum companion mass of \psr{} is plotted as a black arrow.}
\label{fig:pb_mcomp}
\end{figure}

We plotted the relationship in Figure~\ref{fig:pb_mcomp}, together with all binary systems where the companion mass is known, and placed the lower limit of the companion mass of \psr{} in the plot. The figure showed that the binary companion of \psr{} is consistent with that of a He WD. Given the orbital period of \psr{} of $\sim$833\,days, this implies masses of 0.45, 0.47 and 0.49 \Msun\, respectively. The minimum companion mass of 0.41 \Msun\ suggests that the binary is far from edge-on.

Another interesting feature of this type of system is that one can predict their orbital eccentricity. Indeed, for orbital periods larger than 25 days, \cite{phi92} predicts:
\begin{equation} \label{eq:phi}
\left<e^2\right>^{\frac{1}{2}} = 1.5 \times 10^{-4} \frac{P_b}{100\, \rm days},
\end{equation}

\begin{figure}[ht!] 
\includegraphics[width=.95\linewidth]{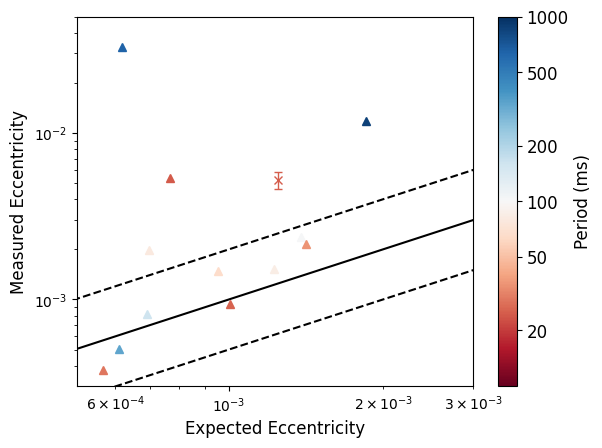} \label{fig:ecc_diff}
\caption{Comparison between the measured eccentricity and the expected eccentricity, as predicted by~\cite{phi92}, of \psr{} (as a cross) and the 12 pulsars likely with WD companion. The solid line is where the measured and expected eccentricity is equal and the dash lines are where the measured and expected eccentricity differs by a factor of 2.}
\label{fig:ecc_diff}
\end{figure}

We plotted the measured eccentricity against the predicted eccentricity from Equation~\ref{eq:phi} for \psr{} and the 12 pulsars likely with WD companions in Figure~\ref{fig:ecc_diff}. Most of the system agreed with the expected value within a factor of 2. However, \psr{} and three other binary systems showed a large deviation from this relationship. Two of these systems, PSRs B0820+02~\citep{mlt+78} and J2208+4610~\citep{dcm+23} have pulsars with spin periods typical of canonical pulsars. PSR B0820+02 shown to have a Carbon-Oxygen WD companion~\citep{kr00}, suggesting a different stellar evolution route where the above relationship will not hold. The companion of PSR J2208+4610 is uncertain, but the system could be similar to that of PSR B0820+02 considering the longer spin period.

Interestingly, \psr{} and its closest counterpart PSR J0214$+$5222 both have measured eccentricities that are 4 and 7 times more than their expected values. This could be a result of variation in the initial condition for the formation of the pulsar-He WD system as noted by  \cite{phi92}, or a result of an extra perturbation on the systems during their evolution. A larger sample of long-period binary pulsar systems will help resolve these possibilities. The FAST telescope has recently discovered 4 new binary pulsars with similarly long binary periods and small eccentricities~\citep{why+25}, with further timing required to model their eccentricities.


\subsection{Optical counterparts} 

There is no counterpart in the GAIA mission catalog~\citep{gaia2016,gaia2023} at a limiting apparent magnitude of $+21$, well below the expected apparent magnitude of a 0.4 M$_\odot$ main sequence star of $m = +18.3$ at a distance of 1 kpc, assuming a luminosity-mass function of $L/\rm L_\odot = (M/\msun)^{3.5}$. However, the limiting magnitude does not preclude the source from possibly being a white dwarf, since most white dwarfs have luminosities that correspond to apparent magnitudes that are below the sensitivity limit of GAIA.

\begin{figure}
    \centering
    \includegraphics[width=0.4\linewidth]{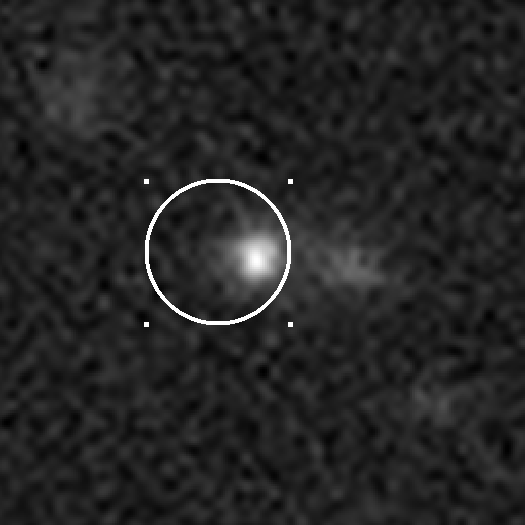}
    \includegraphics[width=0.4\linewidth]{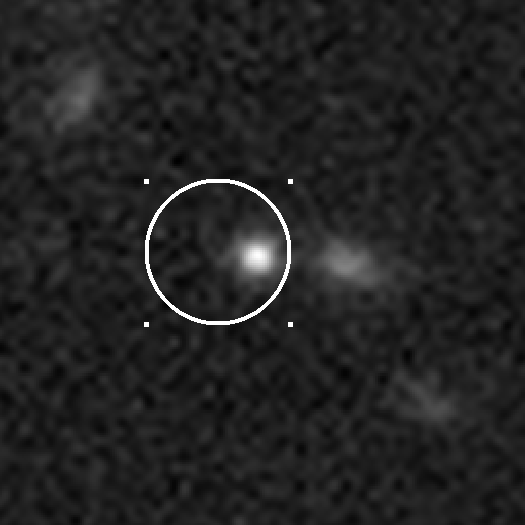}
    \includegraphics[width=0.4\linewidth]{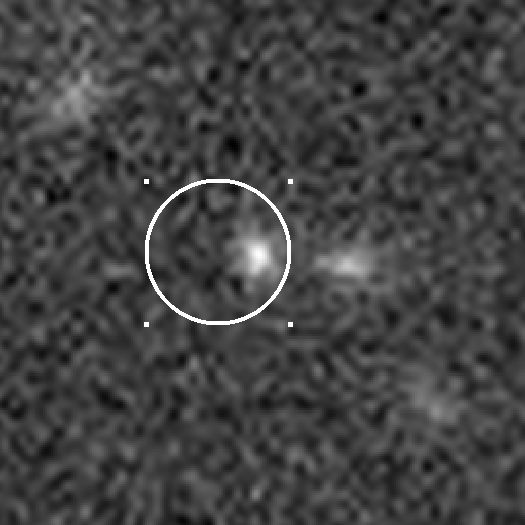}
    \includegraphics[width=0.4\linewidth]{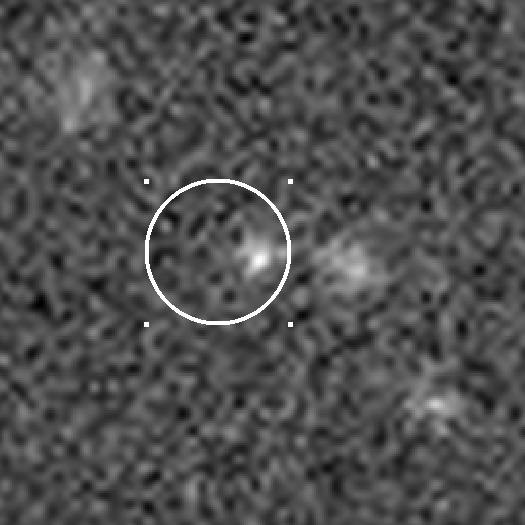}
    \caption{The $g$, $r$, $i$ and $z$ filter $8'' \times 8''$ images centered on the optical source at (R.A., Dec.) $=$ (01h25m47.20s, -58d54m02.35s), as seen by the DECam of the DESI Legacy Surveys. The white circle is the uncertainty region of the position of~\psr{} obtained from the localization efforts of MeerKAT telescope. The source is located on the right side of the uncertainty region of the position of~\psr{}.}
    \label{fig:desi_wd}
\end{figure}

However, data release 10 of the Dark Energy Spectroscopic Instrument (DESI) Legacy Surveys~\citep{dsl+19} contains a $r = 22.03$, $g-r = 0.09$ source at a position of (R.A., Dec.) of (01h25m47.20s, -58d54m02.35s), within the uncertainty region of the position of~\psr{}, as shown in Fig.~\ref{fig:desi_wd}. The optical properties of the source are consistent with those of a white dwarf located at a distance of 0.5-1 kpc. The detection brings in a strong credence that the pulsar is indeed in a binary system with a He WD.

\subsection{Detectability of MSPs in the SKA-Low band}

\psr{} is the first binary MSP discovered by the MWA. This is a significant result, especially considering the modest sensitivity of the SMART and no acceleration searches included in our current pipeline. Indeed, the long binary orbit of the system would mean that the acceleration is negligible in an 80-minute observation. 

As discussed in \S\ref{sec:flux}, we measured a flux density of $17\pm9$\,mJy for \psr{} in the discovery observation (at 154 MHz). This is well above the sensitivity limits of the SMART survey for MSPs ~\citep{bsm+23a}. The paper also describes survey simulations to forecast the MSP yield in full search processing of 80-minute SMART observations. As summarized in \S5.2 of their work, up to 55 MSPs within DMs $\sim$100~\dmu{} are expected from the SMART survey, including up to 15 new discoveries. \psr{} was detected in the first of the 71 observations taken for SMART, which is encouraging and may suggest that we are on track in terms of the expectations for MSPs. However, a caveat in our search pipeline is that we do not perform acceleration searches, which would limit our detectability to only isolated or long period binary MSPs, and thus would result in fewer than 55 MSPs detected.

As mentioned in \S\ref{sec:localisation}, \psr{} is located at a fairly high Galactic latitude of $|b| \approx 57^{\circ}$, which is also the case
with the pulsars discovered in our pilot search~\citep{bsm+23b}, where all four discoveries are located at high Galactic latitudes of $|b| > 15^{\circ}$. It is important to note that this sky region was covered by past surveys, including the 70-cm survey~\citep{mld+96} at a central frequency of 436 MHz, and the High Time Resolution Universe Survey~\citep[HTRU;][]{kjv+10} at a central frequency of 1352 MHz, both with the Parkes \emph{Murriyang} telescope. However, due to the small field of view of Parkes at these frequencies, the surveys limited their dwell times to 157 and 270 seconds, respectively, at high Galactic latitudes. On the other hand, the design of the SMART survey exploits the enormously large field of view of the MWA, enabling much longer (20--40$\times$) dwell times across all parts of the sky. 

While the non-detection of the pulsar in these past surveys may be attributed to their short dwell times at these high latitudes, we note that the 70-cm survey had a limiting sensitivity of $\sim$3 mJy, whereas HTRU reached $\sim$0.3\,mJy at high latitudes. While these limits are nominally higher than our modeled flux densities, which are $\sim$1.5 and $\sim$0.1\,mJy at these two frequencies, it is quite possible that a combination of RFI and unfavorable scintillation may have also contributed to a non-detection. 

On the other hand, a sensitive survey like the ongoing MPIfR-MeerKAT Galactic Plane Survey~\citep[MMGPS;][]{pbs+23}, being conducted using the inner 40 dishes of the MeerKAT telescope, would have detected \psr{} in a 2-minute observation if a similar set up was employed for their UHF or L-band observations. However, the survey currently limits itself to the Galactic plane at $|b| < 11^{\circ}$ due to the limitation of telescope time and the smaller field of view, whereas the SMART survey, while having a modest sensitivity compared to MMGPS, is able to survey the whole southern sky in $<$100 hours of telescope time owing to its large survey speed of $\sim$450\,${\rm deg^2\,h^{-1}}$. The discovery of \psr{} in SMART and its successful follow-ups with MeerKAT also provides an excellent demonstration of how the two major SKA precursors (for SKA-Low and SKA-Mid) can be used effectively to discover new pulsars and conduct their efficient follow-ups.  

The survey strategy we have adopted for SMART draws several parallels with the LOFAR Tied-Array All-Sky Survey~\citep[LOTAAS;][]{scb+19} --
another all-sky pulsar survey at low frequencies (119--151 MHz) with long dwell times (60 minutes), but with a focus in the northern sky. Thus, LOTAAS complements very well with our pulsar searching efforts through the SMART survey, while also providing some overlap in the frequency and sky coverages. As with our current search pipeline, LOTAAS searches also do not include any acceleration searches, yet  of the 73 new pulsars found by the LOTAAS so far, 55 are located at high Galactic latitudes, including two MSPs. Interestingly, both MSPs have long orbital periods on the order of several days. The MSP population to be found by low frequency surveys such as SMART and LOTAAS will thus serve as valuable references that can be used to inform the design of future large pulsar surveys, such as those planned with SKA-Low, and to forecast the prospects of finding new MSPs at low frequencies. 


\section{Summary and Future Outlook} \label{sec:conclusions}

We have discovered \psr{} with a spin period of 24\,ms and a DM of 11.66\,\dmu. It is the first millisecond pulsar discovered by the MWA, and also the first pulsar discovery to result from the ongoing deep-pass searches of SMART. Follow-up observations with the MWA and MeerKAT  have revealed that the pulsar is in a binary system with an orbital period of more than 290 days. While our current data are not sufficient to obtain a fully coherent timing solution, our analysis suggests an orbital period of $833.60 \pm 0.04$ days, a projected semi-major axis of $241.36 \pm 0.05$ light seconds, a minimum companion mass of $ 0.4152 \pm 0.0001 $ \Msun and a low eccentricity of $0.0052 \pm 0.0006$. This indicates that the system is a wide pulsar -- He WD binary. Further timing observations are required to fully understand the nature of \psr{} and its binary companion. 

The discovery of \psr{} underscores the exciting potential of the SMART survey to discover a larger population of MSPs as our search processing ramps up in the coming years. A population analysis that accounts for new pulsar discoveries from the SMART and LOTAAS surveys, and more broadly, an improved understanding of their emission, propagation and spectral properties, will also inform optimal survey strategies and frequency bands for future surveys planned with the SKA-Low, which is expected to take on the bulk of the pulsar survey load in the SKA era.

\begin{acknowledgments}
This scientific work made use of data obtained from Inyarrimanha Ilgari Bundara, the CSIRO Murchison Radio-astronomy Observatory. We acknowledge the Wajarri Yamaji People as the Traditional Owners and Native Title Holders of the observatory site. 
Support for the MWA operations is provided by the Australian Government (NCRIS), under a contract to Curtin University administered by Astronomy Australia Limited (AAL). 
The MeerKAT telescope is operated by the South African Radio Astronomy Observatory, which is a facility of the National Research Foundation, an agency of the Department of Science and Innovation.
Observations used the FBFUSE and APSUSE computing clusters for data acquisition, storage and analysis. These clusters were funded, designed and installed by the Max-Planck-Institut-für-Radioastronomie (MPIfR) and the Max-Planck-Gesellschaft. 
The Parkes \emph{Murriyang} radio telescope is part of the Australia Telescope National Facility which is funded by the Australian Government for operation as a National Facility managed by CSIRO.
CPL and GG were supported by an Australian Government Research Training Program (RTP) Stipend and RTP Fee-Offset Scholarship (\url{https://doi.org/10.82133/C42F-K220}).
VVK acknowledges continuing support from the Max Planck Society and financial support from the European Research Council (ERC) starting grant ``COMPACT" (Grant agreement number 101078094).
We acknowledge the Pawsey Supercomputing Centre which is supported by the Western Australian and Australian Governments.
This work was supported by resources provided by the Pawsey Supercomputing Research Centre’s Setonix Supercomputer (\url{https://doi.org/10.48569/18sb-8s43}), and their Acacia (\url{https://doi.org/10.48569/nfe9-a426}) and Banksia (\url{https://doi.org/10.48569/tnja-4s30}) Object Storage systems.
This work was also supported by resources awarded under AAL's ASTAC merit allocation scheme on the OzSTAR national facility at the Swinburne University of Technology. 
The OzSTAR program receives funding in part from the Astronomy National Collaborative Research Infrastructure Strategy (NCRIS) allocation provided by the Australian Government.

\end{acknowledgments}





%
\facilities{MWA, MeerKAT}

\software{astropy~\citep{2013A&A...558A..33A,2018AJ....156..123A,2022ApJ...935..167A}, {\sc presto}~\citep{rem02,kjr+11}, {\sc dspsr}~\citep{vb11}, {\sc psrchive}~\citep{vdo12}, {\sc peasoup}~\citep{pbs+23}, SeeKAT~\citep{bcb+23}, Mosaic~\citep{cbk+21}, \pulsarspectra{}~\citep{2022PASA...39...56S}, {\sc scintools}~\citep{rcb+20,rc23}, {\sc PSRpy} (\url{https://github.com/emmanuelfonseca/PSRpy}), emcee \citep{fhlg13}, {\sc tempo}~\citep{nds+15}, {\sc dracula}~\citep{2018MNRAS.476.4794F}.}


\appendix

\section{MCMC fit for the orbital parameters of \psr{}} \label{sec:appendix}

We performed MCMC fits using the {\sc emcee} module~\citep{fhlg13} for the orbital parameters of \psr{} by modeling the changes in the observed spin period of \psr{} over MJD 60563--60853 as the motion of the pulsar around a  Keplerian orbit with a single companion. The Keplerian orbit is described with five parameters : The orbital period, $P_b$, the projected semi-major axis, $x$, the argument of periastron, $\omega$, the time of periastron, $T_0$ and the eccentricity of the orbit, $e$. The model also fits for the spin period of the pulsar after removing the effects of the orbital motion, $P_{spin}$. We first obtained an initial guess for the orbital parameter for \psr{}, with $P^i_b = 833.2503192929$ days, $x^i = 239.1245583607$ light seconds, $\omega^i = 0^{\circ}$, $T^i_0 = 60163.7256622941$ MJD and $P^i_{spin} = 24.590574146789940$ ms. We then run MCMC fits on the orbital parameters of the \psr{}, system, with $e$ set to values of $10^{-5}, 10^{-4}, 10^{-3}, 0.01, 0.02, 0.03, 0.05, 0.07, 0.1, 0.2, 0.5, 0.7$. We also restricted the values of the argument of periastron to be $-180^{\circ} \leq \omega \leq 180^{\circ}$ and the time of periastron to be $T^i_{0} - P_b/2 \leq T^i_{0} \leq T^i_{0} + P_b/2$. 

\begin{figure*}[ht!] 
\centering
\includegraphics[width=\linewidth]{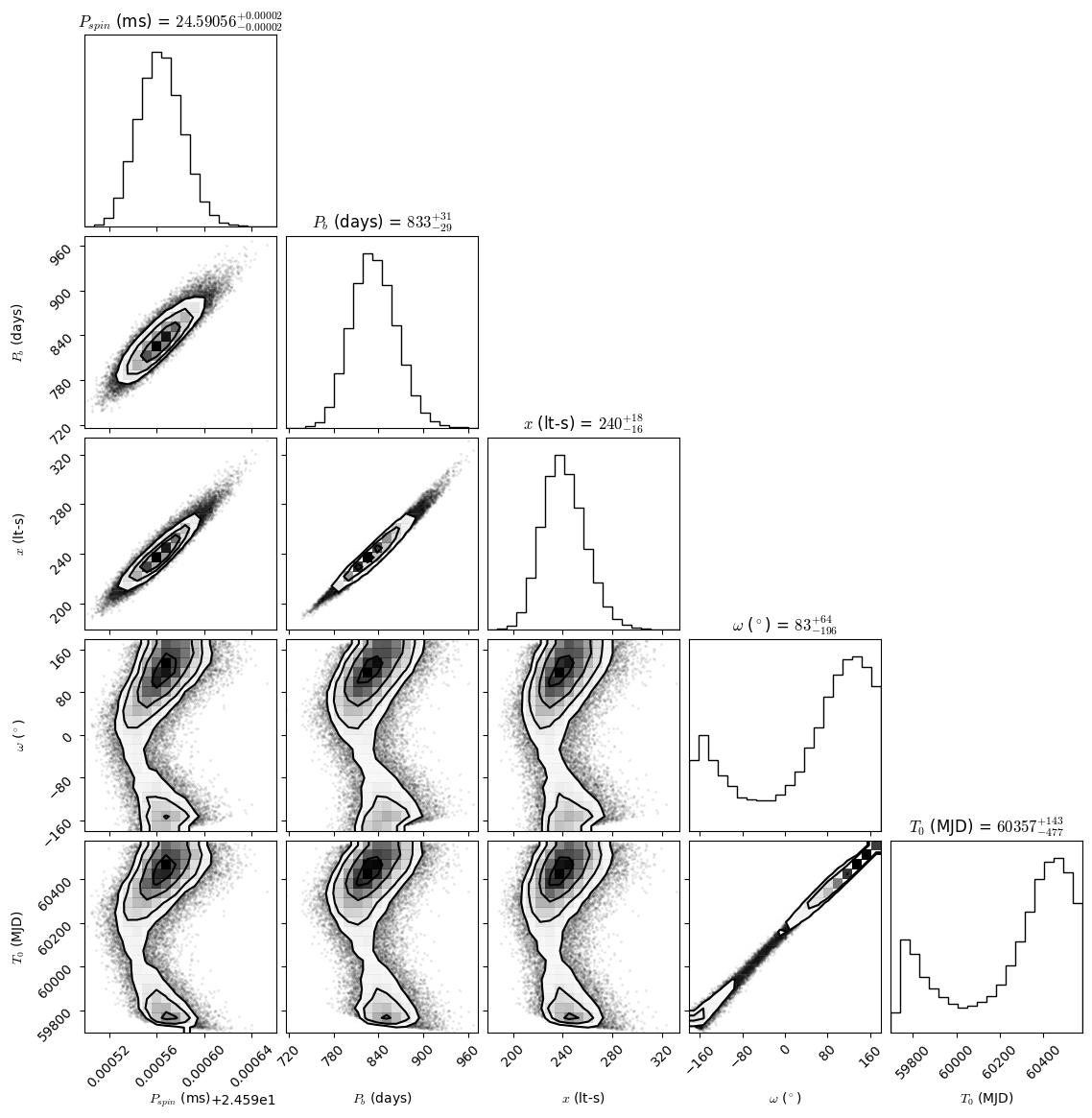}
\caption{Corner plot for MCMC fit for \psr{} orbit, for the case where $e = 0.01$. 
}
\label{fig:mcmc_fit}
\end{figure*}

We were able to obtain good fits for values of $e \leq 0.05$. Figure~\ref{fig:mcmc_fit} shows the corner plot for the MCMC fit for $e = 0.01$, with the 68 per cent confidence values of $P_b =833^{+29}_{-31}$, $x = 240^{+16}_{-18}$, $\omega = 83^{+60}_{-196}$, $T_0 = 60357^{+143}_{-477}$ and $P_{spin} = 24.59056^{+2}_{-2}$. The values for $P_b$, $x$ and $P_{spin}$ are well constrained, while the values of $\omega$ and $T_0$ wrapped around the parameter space due to their cyclic nature. Nevertheless, these constraints allowed us to model the minimum companion mass of the companion to \psr{}, which is found to be 0.41 ± 0.02 M$_{\odot}$. Other values of $e \leq 0.05$ were found to have similar level of constraints in the measured orbital parameters. For cases of $e > 0.05$, we also tried MCMC fits with an initial guess of $P^i_b = 1374.2314446150$ days, $x^i = 569.4536928147$ light seconds, $\omega^i = 0^{\circ}$, $T^i_0 = 61266.6577757726$ MJD and $P^i_{spin} = 24.590931912560528$ ms, but these initial parameters were not able to produce good constraints as well.


\bibliography{sample7}{}
\bibliographystyle{aasjournalv7}



\end{document}